\documentclass[aps,prl,twocolumn,showpacs]{revtex4}
\usepackage{amssymb}
\usepackage{graphicx}
\usepackage{amsmath}
\usepackage{amsfonts}

\begin{document}

\title{$U(1)$ Fermi liquid theory - A Fermi liquid state that supports exclusion statistics}
\date{\today}
\author{Tai-Kai Ng}
\affiliation{Department of Physics, Hong Kong University of Science and
Technology, Clear Water Bay Road, Kowloon, Hong Kong, China}

\begin{abstract}
      We propose in this paper an effective low-energy theory for interacting fermion systems which supports exclusion statistics. The theory can be viewed as an extension of Landau Fermi liquid theory where besides quasi-particle energy $\xi_{\mathbf{k}}$, the kinetic momentum $\mathbf{k}$ of quasi-particles depends also on quasi-particle occupation numbers as a result of momentum ($k$)-dependent current-current interaction. The dependence of kinetic momentum on quasi-particles excitations leads to change in density of states and exclusion statistics. The properties of this new Fermi liquid state is studied where we show that the state (which we call $U(1)$-Fermi liquid state) has Fermi-liquid like properties except that the quasi-particles are {\em not} adiabatically connected to bare fermions in the system and the state may not satisfy Luttinger theorem.

\end{abstract}

\maketitle

\subsection{ I: Introduction}
  The notion of statistical interaction was first proposed by Haldane\cite{ex1}. Let the number of available single particle states for spin $\sigma$ fermions from momentum range $\mathrm{k}$ to $\mathbf{k}+d\mathbf{k}$ be $({dk\over2\pi})^d z_{\sigma}(\mathbf{k})$  where $z_{\sigma}(\mathbf{k})=1$ for normal fermions. Then following Haldane\cite{ex1, ex2}, the exclusion statistics between particles is defined through the relation
      \begin{equation}
     \label{def}
      z_{\sigma}(\mathbf{k})=1-\sum_{\sigma'}\int{d^dk'\over(2\pi)^d}\alpha_{\mathbf{k}\sigma\mathbf{k}'\sigma'} n_{\mathbf{k}'\sigma'}
     \end{equation}
      where $\alpha_{\mathbf{k}\sigma\mathbf{k}'\sigma'}$ is a function determined by the microscopic theory giving rise to exclusion statistics.  It is now understood that exclusion statistics is a general feature of interacting fermion/boson systems described by Bethe-Ansatz solution in one dimension\cite{ex1d}, and also exists in a number of Quantum-Hall type two-dimensional systems with exact known ground state wave-functions\cite{qhex}. Despite the existence of these examples, a general understanding of the microscopic mechanism behind exclusion statistics is still lacking\cite{ex3}.

      In this paper we show how exclusion statistics can arise in a {\em modified} effective Fermi-liquid-type low-energy theory. In usual Fermi liquid theory\cite{landau}, the quasi-particle energy $\xi_{\mathbf{k}}[n]$ are shifted by quasi-particle interaction and external scalar potential with,
      \begin{subequations}
      \label{xi}
      \begin{equation}
      \xi_{\mathbf{k}}[\delta n]=\varepsilon_{\mathbf{k}}+\phi_{\mathbf{k}}[\delta n]+\Phi,
      \end{equation}
      where $\Phi$ is an external scalar potential and
      \begin{equation}
      \label{pk}
       \phi_{\mathbf{k}}={1\over V}\sum_{\mathbf{k}'\sigma'}f_{\mathbf{k}\sigma;\mathbf{k}'\sigma'}\delta n_{\mathbf{k}'\sigma'},
       \end{equation}
       \end{subequations}
      $\varepsilon_{\mathbf{k}}={\delta E\over\delta n_{\mathbf{k}\sigma}}|_{n=n^{(0)}}$ and $f_{\mathbf{k}\sigma;\mathbf{k}'\sigma'}={\delta^2 E\over\delta n_{\mathbf{k}\sigma}\delta n_{\mathbf{k}'\sigma'}}|_{n=n^{(0)}}$ are the usually defined (single-particle) quasi-particle energy and Landau interaction, respectively\cite{landau}. $E[n]$ is the total energy functional and $n_{\mathbf{k}\sigma}=n_{\mathbf{k}\sigma}^{(0)}+\delta n_{\mathbf{k}\sigma}$ is the quasi-particle occupation number for the state $(\mathbf{k}\sigma)$, where $n^{(0)}$ is the ground state quasi-particle occupation number and $\delta n$ is the correction to $n^{(0)}$. In this paper we propose that in addition to the above, the effective interaction between quasi-particles may also leads to {\em shift in the kinetic momentum $\mathbf{k}$ of the quasi-particles} given by
        \begin{subequations}
        \label{km}
        \begin{equation}
        \label{km1}
        \mathbf{k}[n]=\mathbf{k}_0[n^{(0)}]+\mathbf{a}_{\mathbf{k}}[\delta n]+\mathbf{A},
        \end{equation}
         where $\mathbf{A}$ is the external vector (EM) field and
      \begin{equation}
      \label{km2}
      \mathbf{a}_{\mathbf{k}}[\delta n]={1\over V}\sum_{\mathbf{k}'\sigma'}\mathbf{g}_{\mathbf{k}\sigma;\mathbf{k}'\sigma'}\delta n_{\mathbf{k}'\sigma'}
      \end{equation}
      \end{subequations}
       represents an interaction-induced effective $k$-dependent real space gauge field at small $\delta n$. $\mathbf{g}_{\mathbf{k}\sigma;\mathbf{k}'\sigma'}$ is a phenomenological vector function. The microscopic meaning of $\mathbf{a}_{\mathbf{k}}$ and $\mathbf{g}_{\mathbf{k}\sigma;\mathbf{k}'\sigma'}$ will be discussed in section III
       where the Lagrangian and Hamiltonian of the system are introduced. We note that in general a nonzero $\mathbf{a}_{\mathbf{k}}[n^{(0)}]$ may also exists in the ground state which is absorbed in the definition of $\mathbf{k}_0[n^{(0)}]$. We shall assume here a spin-isotropic system such that $\varepsilon_{k}$, $\phi_{\mathbf{k}}$ and $\mathbf{a}_{\mathbf{k}}$ are independent of spin $\sigma$. We shall call the resulting fermion liquid state a $U(1)$ Fermi liquid state since the interaction induced $\phi_{\mathbf{k}}$ and $\mathbf{a}_{\mathbf{k}}$ fields are the scalar and vector components of an effective $k$-dependent $U(1)$ gauge field acting on the fermions, respectively.

\subsection{II: Dynamical properties of $U(1)$ Fermi liquid state}

     Specifically we consider a fermion system in external EM fields $\Phi$ and $\mathbf{A}$ with an effective low energy Hamiltonian
      \begin{eqnarray}
      \label{hgauge}
       E[\delta n_{\mathbf{k}}, \mathbf{A}_{\mathbf{k}}]
      & = & \sum_{\mathbf{k}_0\sigma}\varepsilon_{\mathbf{k}}\delta n_{\mathbf{k}\sigma}+\Phi\sum_{\mathbf{k}_0\sigma}\delta n_{\mathbf{k}\sigma} \\ \nonumber
   %   & & +{1\over2V}\sum_{\mathbf{k}_0\mathbf{k}'_0}\mathbf{A}_{\mathbf{k}}\mathbf{\Gamma}^{-1}_{\mathbf{k}\mathbf{k}'}\mathbf{A}_{\mathbf{k}'}
   %   \\ \nonumber
      & & +{1\over2V}\sum_{\mathbf{k}_0\sigma,\mathbf{k}_0'\sigma'}
      f_{\mathbf{k}\sigma,\mathbf{k}'\sigma'}\delta n_{\mathbf{k}\sigma}\delta n_{\mathbf{k}'\sigma'}.
      \end{eqnarray}
    where $\mathbf{k}(\mathbf{r})=\mathbf{k}_0+\mathbf{A}_{\mathbf{k}}(\mathbf{r})$, $\mathbf{A}_{\mathbf{k}}=\mathbf{a}_{\mathbf{k}}+\mathbf{A}$ and the other terms have their usual meaning. $E[\delta n_{\mathbf{k}}, \mathbf{A}_{\mathbf{k}}]$ describes a Landau-Fermi liquid Hamiltonian\cite{landau} except that the quasi-particle kinetic momenta are {\em shifted} in the presence of ($k$-dependent) real-space gauge field $\mathbf{A}_{\mathbf{k}}$.  We shall assume that $\mathbf{a}_{\mathbf{k}}$ is given by Eq.\ (\ref{km2}) and examine its consequences in this section.

       The relation between the $\mathbf{k}$-dependent effective gauge field $\mathbf{A}_{\mathbf{k}}$ and exclusion statistics can be seen most easily by noting that the shift in kinetic momentum $\mathbf{k}_0\rightarrow\mathbf{k}$ results in a coordinate transformation of the system when the quasi-particle occupation numbers $n_{\mathbf{k}\sigma}(\mathbf{r})$ are characterized by $\mathbf{k}$ and $\mathbf{r}$, where
       \begin{equation}
       \label{trans}
       {1\over V}\sum_{\mathbf{k}_0}=\int {d^dk_0\over(2\pi)^d}\rightarrow\int z(\mathbf{k}){d^dk\over(2\pi)^d}
       \end{equation}
       where $z(\mathbf{k})=(Det|\mathbf{M}(\mathbf{k})|)$ is the Jacobian for the transformation between the coordinates $\mathbf{k}_0$ and $\mathbf{k}$,
       $\mathbf{M}_{ij}(\mathbf{k})={\partial k_{0i}\over\partial k_j}=\delta_{ij}-\partial_{k_{j}}A_{\mathbf{k}i}$ is a $d\times d$ matrix. The Jacobian $z(\mathbf{k})$ describes a renormalization in the density of states (DOS) of the system when the phase space is transformed from canonical- to kinetic- momentum representation\cite{Niu}. The dependence of $z(\mathbf{k})$ on $\delta n$ is precisely exclusion statistics (see Eq.\ (\ref{def}) and section IV).

   \subsubsection{low energy dynamics}
  The effective low energy theory we describe here is similar to the Landau-Silin theory\cite{ls} for charged fermions in external gauge fields $\Phi$ and  $\mathbf{A}$ except that the effective gauge field $\mathbf{A}_{\mathbf{k}}$ is momentum ($k$)-dependent. The dynamics of the quasi-particles are given by the usual Hamilton equations of motion
    \begin{eqnarray}
     \dot{\mathbf{r}} & = & \nabla_{\mathbf{k}_0}\xi_{\mathbf{k}}[\delta n];  \\ \nonumber
     \dot{\mathbf{k}}_0 & = & -\nabla_{\mathbf{r}}\xi_{\mathbf{k}}[\delta n]
     \end{eqnarray}
     where $\xi_{\mathbf{k}}[n]$ is given by Eq.\ (\ref{xi}). We obtain after some straightforward algebra
      \begin{eqnarray}
     \label{vel}
     \dot{\mathbf{r}}_{\mu} & = & \sum_{\nu}({\partial k_{\nu}\over\partial_{k_{0\mu}}})\partial_{k_{\nu}}\xi_{\mathbf{k}}[\delta n(\mathbf{r})] \\ \nonumber
     & = & \sum_{\nu}\left[\mathbf{M}(\mathbf{k})^{-1}\right]_{\nu\mu}\left(v_{\mathbf{k}\nu}+\sum_{\mathbf{k}'\sigma'}\partial_{k_{\nu}}f_{\mathbf{k}\sigma;\mathbf{k}'\sigma'}\delta n_{\mathbf{k}'\sigma'}(\mathbf{r})\right),
     \end{eqnarray}
      and
      \begin{eqnarray}
      \label{force}
       \dot{\mathbf{k}} & = &
       =-\nabla_{\mathbf{r}}\xi_{\mathbf{k}}[\delta n]+\dot{\mathbf{A}}_{\mathbf{k}} \\ \nonumber
       & = & -\sum_{\mathbf{k}'\sigma'}f_{\mathbf{k}\sigma;\mathbf{k}'\sigma'}\nabla_{\mathbf{r}}\delta n_{\mathbf{k}'\sigma'}(\mathbf{r})-\nabla_\mathbf{r}\Phi  \\ \nonumber
       & & +{\partial \mathbf{A}_{\mathbf{k}}\over\partial t}-\dot{\mathbf{r}}\times\mathbf{B}_{\mathbf{k}}(\mathbf{r})+(\dot{\mathbf{k}}.\nabla_{\mathbf{k}})\mathbf{A}_{\mathbf{k}}
     \end{eqnarray}
     to leading order in $\Phi$ and $\delta n$, where $\mathbf{v}_{\mathbf{k}}=\nabla_{\mathbf{k}}\varepsilon_{\mathbf{k}}$ and $\mathbf{B}_{\mathbf{k}}=\nabla_{\mathbf{r}}\times\mathbf{A}_{\mathbf{k}}$.
     Eqs.\ (\ref{vel}) and\ (\ref{force}) are the same as the Landau-Silin equations\cite{ls} for charged quasi-particles in external EM field except the presence of $[\mathbf{M}(\mathbf{k})]^{-1}$, $(\dot{\mathbf{k}}.\nabla_{\mathbf{k}})\mathbf{A}_{\mathbf{k}}$ and $\nabla_{\mathbf{k}}.\mathbf{A}_{\mathbf{k}}$ terms coming from the $k$-dependence of the gauge field $\mathbf{A}_{\mathbf{k}}$. In particular,
     \begin{eqnarray}
     \label{emeff}
     \dot{\mathbf{r}} & \rightarrow & \mathbf{v}_{\mathbf{k}},  \\ \nonumber
     \dot{\mathbf{k}} & \rightarrow & -\sum_{\mathbf{k}'\sigma'}f_{\mathbf{k}\sigma;\mathbf{k}'\sigma'}\nabla_{\mathbf{r}}\delta n_{\mathbf{k}'\sigma'}(\mathbf{r})-\nabla_\mathbf{r}\Phi+{\partial \mathbf{A}_{\mathbf{k}}\over\partial t}
     \end{eqnarray}
     in the linearized Boltzmann equation
     \begin{equation}
      \label{beq1}
      {d n_{\mathbf{k}\sigma}\over dt}={\partial n_{\mathbf{k}\sigma}\over\partial t}+\dot{\mathbf{r}}.\nabla_{\mathbf{r}}n_{\mathbf{k}\sigma}+
       \dot{\mathbf{k}}.\nabla_{\mathbf{k}}n_{\mathbf{k}\sigma}=\left({\partial n_{\mathbf{k}\sigma}\over\partial t}\right)_{col},
      \end{equation}
       and $\mathbf{A}_{\mathbf{k}}$ introduces only an effective $k$-dependent electric field on the system.

   \subsubsection{quasi-particles charge and current}
      Using Eq.\ (\ref{trans})), the charge carried by quasi-particles is given in the linear response regime by
     \[
      Q(\mathbf{r},t)=\sum_{\sigma}\int {d^dk_0\over(2\pi)^d}n_{\mathbf{k}\sigma}(\mathbf{r},t)=\sum_{\sigma}\int {d^dk\over(2\pi)^d}z(\mathbf{k})n_{\mathbf{k}\sigma}(\mathbf{r},t)  \]
      It is straightforward to show that
      \begin{equation}
      \label{zk}
      z(\mathbf{k})=1-\nabla_{\mathbf{k}}\cdot\mathbf{A}_{\mathbf{k}}+O(\mathbf{A}_{\mathbf{k}})^2
      \end{equation}
      and we obtain in the linear response regime
     \begin{eqnarray}
      \label{qeff}
          Q(\mathbf{r},t) & = & \sum_{\sigma}\int {d^dk\over(2\pi)^d}\left(\delta n_{\mathbf{k}\sigma}(\mathbf{r},t)-(\nabla_{\mathbf{k}}\cdot\mathbf{A}_{\mathbf{k}}(\mathbf{r}))n^{(0)}_{\mathbf{k}}\right)  \\ \nonumber
       & = & \sum_{\sigma}\int {d^dk\over(2\pi)^d}e^*_{\mathbf{k}\sigma}\delta n_{\mathbf{k}\sigma}(\mathbf{r},t)
          \end{eqnarray}
    where
       \begin{eqnarray}
       \label{echarge}
       e^*_{\mathbf{k}\sigma} & = & 1-\sum_{\sigma'}\int{d^dk'\over(2\pi)^d} n^{(0)}_{\mathbf{k}'}(\nabla_{\mathbf{k}'}\cdot\mathbf{g}_{\mathbf{k}'\sigma'\mathbf{k}\sigma})  \\ \nonumber
       & = & 1+\sum_{\sigma'}\int{d^dk'\over(2\pi)^d}(\nabla_{\mathbf{k}'}n^{(0)}_{\mathbf{k}'})\cdot\mathbf{g}_{\mathbf{k}'\sigma'\mathbf{k}\sigma}
       \end{eqnarray}
       is the effective charge carried by the quasi-particles. We note that in general $e^*\neq1$ as long as $\nabla_{\mathbf{k}'}\cdot\mathbf{g}_{\mathbf{k}'\sigma'\mathbf{k}\sigma}\neq0$, indicating that the quasi-particles in $U(1)$-Fermi liquids are not adiabatically connected to the "bare" particles. In particular, the Fermi sea volume in $U(1)$ Fermi liquid may not satisfy the Luttinger Theorem\cite{lut} and the $U(1)$ Fermi liquid state is not a Landau Fermi liquid state. We note also that although exclusion statistics is in general a bulk effect that affects the whole Fermi sea, it can be expressed in terms of properties on the Fermi surface alone at zero temperature when $\nabla_{\mathbf{k}'}n^{(0)}_{\mathbf{k}'}\sim-\delta(\xi_{\mathbf{k}})$.

       We next consider the current carried by quasi-particles.
       We start by examining the equation of motion of the charge $Q(\mathbf{r},t)$. Putting together Eq.\ (\ref{emeff}), \ (\ref{beq1}) and\ (\ref{qeff}), we obtain in the linear response regime,
       \begin{eqnarray}
       {\partial Q\over\partial t} & = & \sum_{\sigma}\int{d^dk\over(2\pi)^d}{\partial\over\partial t}\left(\delta n_{\mathbf{k}\sigma}(\mathbf{r},t)+\mathbf{A}_{\mathbf{k}}(\mathbf{r})).\nabla_{\mathbf{k}}n^{(0)}_{\mathbf{k}}\right)  \\ \nonumber
       & = & -\sum_{\sigma}\int{d^dk\over(2\pi)^d}\left(\mathbf{v}_{\mathbf{k}}.\nabla_{\mathbf{r}}\delta n_{\mathbf{k}\sigma}+
       (\dot{\mathbf{k}}-{\partial\mathbf{A}_{\mathbf{k}}\over\partial t}).\nabla_{\mathbf{k}}n_{\mathbf{k}\sigma}^{(0)}\right)  \\ \nonumber
       & = & -\nabla_{\mathbf{r}}.\sum_{\sigma}\int{d^dk\over(2\pi)^d}\mathbf{j}_{\mathbf{k}}\delta n_{\mathbf{k}\sigma}
       \end{eqnarray}
       where
       \begin{equation}
       \label{current}
       \mathbf{j}_{\mathbf{k}}=\mathbf{v}_{\mathbf{k}}-\sum_{\mathbf{k}'\sigma'}\mathbf{v}_{\mathbf{k}'}{\partial n^{(0)}(\varepsilon')\over\partial\varepsilon'}|_{\varepsilon'=\mu}f_{\mathbf{k}'\sigma'\mathbf{k}\sigma}.
      \end{equation}
      is the current carried by quasi-particles as in usual Landau Fermi liquid theory. Notice that the charge and current of quasi-particles are renormalized separately by the space- and time components of effective gauge field $\mathbf{a}_{\mathbf{k}}$ and $\phi_{\mathbf{k}}$, respectively in $U(1)$ Fermi liquid.

     \subsubsection{ solution of the transports equation}
       We now consider the general solution of the Boltzmann transport equation in the collisionless regime $\omega\tau>>1$ where the collision term can be neglected. We first consider the system in an external scalar potential $\Phi(\mathbf{r},t)\sim\Phi_0e^{i(\mathbf{q}.
       \mathbf{r}-\omega t)}$. In this case$\delta n_{\mathbf{k}\sigma}(\mathbf{r},t)\sim\delta n_{\mathbf{k}}e^{i(\mathbf{q}.\mathbf{r}-\omega t)}$ is spin-independent. Putting these into the transport equations\ (\ref{emeff}) and \ (\ref{beq1}), we obtain
       \begin{eqnarray}
       \label{sol1}
       & & (\omega -\mathbf{q}\cdot \mathbf{v}_{\mathbf{k}})\delta n_{\mathbf{k}}+\frac{\partial n_{\mathbf{k}}^{0}}{\partial \varepsilon _{\mathbf{k}}}\left[(\mathbf{q}\cdot \mathbf{v}_{\mathbf{k}})\left(\Phi_0+\sum_{\mathbf{k}^{\prime }}f_{\mathbf{k}\mathbf{k}'}^{s}\delta n_{\mathbf{k}'}\right)\right.  \\ \nonumber
       & & \left.+\omega\mathbf{v}_{\mathbf{k}}.\sum_{\mathbf{k}^{\prime }}\mathbf{g}_{\mathbf{k}\mathbf{k}'}^{s}\delta n_{\mathbf{k}'}\right]=0,
       \end{eqnarray}
       where $f(\mathbf{g})_{\mathbf{k}\mathbf{k}'}^{s}=f(\mathbf{g})_{\mathbf{k}\uparrow\mathbf{k}'\uparrow} +f(\mathbf{g})_{\mathbf{k}\uparrow\mathbf{k}'\downarrow}$. Rearranging terms, the equation can be rewritten as
        \begin{eqnarray}
       \label{sol2}
       & & (\omega -\mathbf{q}\cdot \mathbf{v}_{\mathbf{k}})\left(\delta n_{\mathbf{k}}+\mathbf{v}_{\mathbf{k}}.\frac{\partial n_{\mathbf{k}}^{0}}{\partial \varepsilon _{\mathbf{k}}}\sum_{\mathbf{k}'}\mathbf{g}_{\mathbf{k}\mathbf{k}'}^{(s)}\delta n_{\mathbf{k}'}\right)  \\ \nonumber
       & & =-(\mathbf{q}\cdot \mathbf{v}_{\mathbf{k}})\frac{\partial n_{\mathbf{k}}^{0}}{\partial \varepsilon _{\mathbf{k}}}\left[\Phi_0+\sum_{\mathbf{k}'}\left(f_{\mathbf{k}\mathbf{k}'}^{s}+\mathbf{v}_{\mathbf{k}}\cdot \mathbf{g}_{\mathbf{k}\mathbf{k}'}^{s}\right)\delta n_{\mathbf{k}'}\right].
       \end{eqnarray}

      It is clear that we can write $\delta n_{\mathbf{k}}=-{\partial n_{\mathbf{k}}^{0}\over\partial \varepsilon_{\mathbf{k}}}\nu_{\mathbf{k}}$. For isotropic systems in three dimensions, we can expand\cite{landau, Pethick}
       \begin{eqnarray}
       \label{Fs}
        f_{\mathbf{k}\mathbf{k}'}^{s(a)} & = & \sum_{l=0}^{\infty}f_{l}^{s(a)}P_{l}(\cos \theta_{\mathbf{k}\mathbf{k}'} ) \\ \nonumber
        \mathbf{v}_{\mathbf{k}}\cdot\mathbf{g}_{\mathbf{k}\mathbf{k}'}^{(s)} & = & \sum_{l=0}^{\infty}g_{l}^{s(a)}P_{l}(\cos \theta_{\mathbf{k}\mathbf{k}'} )  \\ \nonumber
        \nu _{\mathbf{k}} & = & \sum_{l}P_{l}(\cos \theta _{\mathbf{k}\mathbf{q}})\nu _{l}
        \end{eqnarray}
       where $P_{l}$'s are Legendre polynomials and $\theta _{\mathbf{k}\mathbf{p}}$ denotes the angle between $\mathbf{k}$ and $\mathbf{p}$.
 Eq.\ (\ref{sol2}) can be simplified using the properties of spherical harmonics. We obtain after some lengthy but standard algebra\cite{Pethick},
  \begin{equation}
  \label{sol3}
\frac{\nu _{l}}{2l+1}\left(1-\frac{G_{l^{\prime }}^{s}}{2l+1}\right)+\sum_{l^{\prime }}\tilde{F}_{l^{\prime }}^{s}\Omega
_{ll^{\prime }}(s)\frac{\nu _{l^{\prime }}}{2l^{\prime }+1}=-\Omega_{l0}(s)\Phi_0,
\end{equation}%
where $s=\omega/qv_F$, $G(F)_{l}^{s}=N(0)g(f)_{l}^{s}$, $\tilde{F}_{l}^{s}=F_{l}^{s}+G_l^s$, $N(0)$ is the DOS on Fermi surface and
\begin{equation*}
\Omega _{ll^{\prime }}(s)=\frac{1}{2}\int_{-1}^{1}d\mu P_{l}(\mu )\frac{\mu
}{\mu -s}P_{l^{\prime }}(\mu ).
\end{equation*}

  Using eqs.\ (\ref{echarge}) and\ (\ref{current}) we also obtain for the physical charge and current,
\begin{eqnarray}
\label{c&j}
Q & = & (1-G^s_0)N(0)\nu_0  \\ \nonumber
J & = & (1+{F^s_1\over3})N(0){\nu_1\over3}.
\end{eqnarray}
The second equation (current renormalization) is the standard Landau Fermi liquid result\cite{landau, Pethick}. The first equation (charge renormalization) is a new property of the $U(1)$ Fermi liquid state.

 The density-density response of the system $\chi_d$ can be obtained easily from Eq.\ (\ref{sol3}) if we keep only $l=0,1$ components. We obtain after some straightforward algebra,
 \begin{equation}
 \label{dd}
 \chi_d(s)={\chi_0(s)\over1-\left({\tilde{f}_0^s\over(1-G_0^s)}+({\tilde{f}_1^s\over1+{F_1^s/3}})s^2\right)\chi_0(s)}
 \end{equation}
 where $\chi_0(s)$ is the Lindhard function.

   The (transverse) conductivity $\sigma_(s)$ can be obtained similarly with $-\mathbf{v}_{\mathbf{k}}\cdot\mathbf{q}\Phi\rightarrow\omega\mathbf{v}_{\mathbf{k}}\cdot\mathbf{E}$ and $\mathbf{E}.\mathbf{q}=0$, where $\mathbf{E}=i\omega\mathbf{A}$ is the electric field and $\theta_{\mathbf{k}\mathbf{E}}$ is the angle between $\mathbf{k}$ and $\mathbf{E}$. Expanding $\nu _{\mathbf{k}}=\sum_{l}P_{l}(\cos \theta _{\mathbf{k}\mathbf{E}})\nu _{l}$, we obtain after some similar algebra,
   $\mathbf{J}=\sigma(s)\mathbf{E}$ where
   \begin{equation}
 \label{dd}
 \sigma(s)={\sigma_{0}(s)\over1-({\tilde{f}_1^s\over1+{F_1^s/3}})(i\omega)\sigma_{0}(s)},
 \end{equation}
  $\sigma_{0}(s)$ is the conductivity for non-interacting fermions. We note that the (extended)-Landau interaction terms $f^s$ and $g^s$ together produce effective interactions between charges and currents. However they play different roles in the renormalization of the quasi-particle charge and current.

\subsection{III: Lagrangian and Hamiltonian formulation}
   The transports equation we study in section II is supported by a proper Lagrangian and Hamiltonian formulation we derive here.
   For simplicity we shall consider a spinless system in this section. The extension to the spin-ful case is straightforward.

    We start with the following phenomenological Lagrangian\cite{Niu,Sun}
    \begin{eqnarray}
    \label{l1}
    L[n,\mathbf{A},\Phi] & = & L_G+ L_1[\delta n;\mathbf{a},\phi]+ L_2[\mathbf{a},\phi] \\ \nonumber
      L_1 & = & -\sum_{\mathbf{k}}\delta n_{\mathbf{k}} \left(\dot{\mathbf{k}}.\mathbf{r}_{\mathbf{k}}+\dot{r}_{\mathbf{k}}.\mathbf{a}_{\mathbf{k}}+\varepsilon_{\mathbf{k}}
    +\phi_{\mathbf{k}}\right)  \\ \nonumber
     L_2 & = & {1\over 2V}\sum_{\mathbf{k};\mathbf{k}'} \left(\sum_{\mu\nu=1,..d}a_{\mathbf{k}\mu}[\mathbf{G}^{-1}]_{\mathbf{k}\mu\mathbf{k}'\nu}a_{\mathbf{k}'\nu}\right. \\ \nonumber
    & & \left.+\phi_{\mathbf{k}}[\mathbf{f}^{-1}]_{\mathbf{k}\mathbf{k}'}\phi_{\mathbf{k}'}\right)
    \end{eqnarray}
   where $L_G$ is the Lagrangian describing the ground state and $L_1$ describes the low energy dynamics of fermions in effective gauge fields $\phi_{\mathbf{k}}$ and $\mathbf{a}_{\mathbf{k}}$. $\mathbf{r}_{\mathbf{k}}$ is the position of the particle with momentum $\mathbf{k}$ and kinetic energy $\varepsilon_{\mathbf{k}}$. $ L_2$ provides the dynamics for the effective gauge fields.

     The Euler-Lagrange equations ${\delta L\over\delta\mathbf{k}}-{d\over dt}{\delta L\over\delta\dot{\mathbf{k}}}=0$,
     ${\delta L\over\delta\mathbf{r}}-{d\over dt}{\delta L\over\delta\dot{\mathbf{r}}}=0$ and ${\delta L\over\delta\mathbf{a}_{\mathbf{k}}}=0$,  ${\delta L\over\delta\phi_{\mathbf{k}}}=0$ give rise to the equation of motions\ (\ref{vel}) and\ (\ref{force}) in the main text with
     \begin{subequations}
     \label{lag1}
     \begin{equation}
     \label{lag2}
     \phi_{\mathbf{k}}={1\over V}\sum_{\mathbf{k}'}f_{\mathbf{k}\mathbf{k}'}\delta n_{\mathbf{k}'}
     \end{equation}
     and
     \begin{equation}
     \label{lag3}
     \mathbf{a}_{\mathbf{k}}={1\over V}\sum_{\mathbf{k}'}\mathbf{G}_{\mathbf{k}\mathbf{k}'}\cdot\dot{\mathbf{r}}_{\mathbf{k}'}\delta n_{\mathbf{k}'}\sim{1\over V}\sum_{\mathbf{k}'}\mathbf{G}_{\mathbf{k}\mathbf{k}'}\cdot\mathbf{v}_{\mathbf{k}'}\delta n_{\mathbf{k}'}
     \end{equation}
     \end{subequations}
    to leading order in $\delta n$, which are just Eqs.\ (\ref{pk}) and\ (\ref{km2}) if we identify $f_{\mathbf{k}\mathbf{k}'}$
    as the Landau parameters and $\mathbf{G}_{\mathbf{k}\mathbf{k}'}\cdot\mathbf{v}_{\mathbf{k}'}=\mathbf{g}_{\mathbf{k}\mathbf{k}'}$ (with spin neglected), confirming that $L$ is the Lagrangian for the $U(1)$ Fermi liquid state.

      Eliminating $\phi_{\mathbf{k}}$ and $\mathbf{a}_{\mathbf{k}}$ directly from Eq,\ (\ref{l1}), we also obtain
      \begin{subequations}
      \label{l2}
      \begin{equation}
    L[n,\mathbf{A},\Phi] \rightarrow L_G+ L_0[\delta n]+ L_I[\delta n]
    \end{equation}
    where
    \begin{eqnarray}
     \label{l22}
     L_0 & = & \sum_{\mathbf{k}\sigma}\delta n_{\mathbf{k}} \left(-\dot{\mathbf{k}}.\mathbf{r}_{\mathbf{k}}-\varepsilon_{\mathbf{k}}\right)  \\ \nonumber
     L_I & = & -{1\over 2V}\sum_{\mathbf{k};\mathbf{k}'}\delta n_{\mathbf{k}}\left[f_{\mathbf{k}\mathbf{k}'}+
    \dot{\mathbf{r}}_{\mathbf{k}}\cdot\mathbf{G}_{\mathbf{k}\mathbf{k}'}\cdot\dot{\mathbf{r}}_{\mathbf{k}'}\right]\delta n_{\mathbf{k}'}.
    \end{eqnarray}
    \end{subequations}
     It is obvious that $ L_0$ describes the single particle dynamics of quasi-particles in $U(1)$ Fermi liquid theory and $L_2$ describes interaction between quasi-particles. In particular, {\em the $k$-dependent vector potential $\mathbf{a}_{\mathbf{k}}$ is originated from current-current interaction between quasi-particles}. The corresponding Hamiltonian of the system is
     \begin{equation}
     \label{h1}
     H[\delta n, \mathbf{a}_{\mathbf{k}}]=E_G+E[\delta n_{\mathbf{k}},\mathbf{a}_{\mathbf{k}}]-{1\over 2V}\sum_{\mathbf{k}\mathbf{k}'\mu,\nu} a_{\mathbf{k}\mu}[\mathbf{G}^{-1}]_{\mathbf{k}\mu\mathbf{k}'\nu}a_{\mathbf{k}'\nu}.
     \end{equation}
     where $E_G$ is the ground state energy for the $U(1)$ Fermi liquid. Notice that $\delta H/\delta \mathbf{a}_{\mathbf{k}}=0$ leads to Eq.\ (\ref{lag3}) for $\mathbf{a}_{\mathbf{k}}$.

\subsubsection{Thermodynamics}
   The low temperature thermodynamics of the system can be obtained by expanding $H[\delta n, \mathbf{a}_{\mathbf{k}}]$ to second order in $\delta n_{\mathbf{k}}$. Using
   \[
   \varepsilon_{\mathbf{k}}\delta n_{\mathbf{k}}\sim(\varepsilon_{\mathbf{k}_0}+\mathbf{a}_{\mathbf{k}}\cdot\nabla_{\mathbf{k}_0}\varepsilon_{\mathbf{k}_0})\delta n_{\mathbf{k}_0} \]
   to second order in $\delta n$, we obtain
   \begin{eqnarray}
    \label{h2}
    H & \rightarrow & E_G+\sum_{\mathbf{k}_0}\varepsilon_{\mathbf{k}_0}\delta n_{\mathbf{k}_0}  \\ \nonumber
    & & +{1\over 2V}\sum_{\mathbf{k}_0;\mathbf{k}_0'}\delta n_{\mathbf{k}_0}\left[f_{\mathbf{k}_0\mathbf{k}_0'}+
    \mathbf{v}_{\mathbf{k}_0}\cdot\mathbf{G}_{\mathbf{k}_0\mathbf{k}_0'}\cdot\mathbf{v}_{\mathbf{k}_0'}\right]\delta n_{\mathbf{k}_0'} \\ \nonumber
    & = & \sum_{\mathbf{k}_0}\varepsilon_{\mathbf{k}_0}\delta n_{\mathbf{k}_0}+{1\over 2V}\sum_{\mathbf{k}_0;\mathbf{k}_0'}\delta n_{\mathbf{k}_0}\tilde{f}_{\mathbf{k}_0\mathbf{k}_0'}\delta n_{\mathbf{k}_0'},
  \end{eqnarray}
    which is a standard Fermi liquid energy functional with effective Landau interaction $\tilde{f}_{\mathbf{k}\mathbf{k}'}$. This result suggests that the low energy thermodynamical behaviors of $U(1)$ Fermi liquid states are the same usual Fermi liquid states\cite{landau, Pethick}. In particular, a linear specific heat $C_V=\gamma T$ ($T=$ temperature) is expected when the system is stable, i.e. when
    \[ 1+{\tilde{F}_l\over2l+1}>0, \]
    for all $l$. The only difference between $U(1)$ Fermi liquid and usual Fermi liquid is that the Landau interaction $f_{\mathbf{k}\mathbf{k}'}$ is replaced by the extended Landau interaction $\tilde{f}_{\mathbf{k}\mathbf{k}'}$.

\subsection{IV: Summary and Discussion}
         In this paper we study a class of fermion liquids described by a modified Landau Fermi liquid type low energy theory where besides quasi-particle energy, the quasi-particle kinetic momentum is also renormalized by interaction between quasi-particles. We show that the resulting $U(1)$ Fermi liquid state exhibits exclusion statistics and the quasi-particles are {\em not} adiabatically connected to the non-interacting fermion states as a result. The renormalization of kinetic momentum is a consequence of momentum-dependent current-current interaction between quasi-particles.

        The dynamics of the $U(1)$ Fermi liquid state is studied where we show that the resulting density-density and current-current response functions are Fermi-liquid like, except the appearance of `extended' Landau parameters which renormalize not only current but also charge in the Fermi liquid state. This result is rather non-trivial since exclusion statistics is in general a bulk effect that affects the whole Fermi sea whereas only quasi-particle states on the Fermi surface are important in Fermi liquid theory. We find that the effects of exclusion statistics can be expressed in terms of Fermi surface properties at zero temperature, when $\nabla_{\mathbf{k}}n^{(0)}_{\mathbf{k}}\sim-\delta(\xi_{\mathbf{k}})$ (Eq.\ (\ref{echarge})).

        Thermodynamically the $U(1)$ Fermi liquid is also Fermi-liquid like except that the Fermi sea volume may not satisfy the Luttinger Theorem and the Landau interaction $f_{\mathbf{k}\mathbf{k}'}$ is replaced by the extended Landau interaction $\tilde{f}_{\mathbf{k}\mathbf{k}'}$. The $U(1)$ Fermi liquid state is an example of `marginal' Fermi liquid state where the low temperature properties of the system are Fermi liquid like except that the quasi-particles are {\em not} adiabatically connected to non-interacting fermions. This is perhaps not too surprising since we know that in one dimension, exclusion statistics is consistent with effective Luttinger-liquid-like low energy theories\cite{ex2}.

   \subsubsection{microscopic origin of exclusion statistics}

   The factor $z(\mathbf{k})$ describes a modification of the DOS as a result of transformation from canonical coordinates $(\mathbf{r},\mathbf{k}_0)$ to kinetic coordinates $(\mathbf{r},\mathbf{k})$.
   Comparing Eq.\ (\ref{zk}) with Eq.\ (\ref{def}) we obtain an excitation-driven exclusion statistics effect with
   $n_{\mathbf{k}\sigma}\rightarrow \delta n_{\mathbf{k}\sigma}$ and
   \[
   \alpha_{\mathbf{k}\sigma\mathbf{k}'\sigma'}\rightarrow \nabla_{\mathbf{k}}.\mathbf{g}_{\mathbf{k}\sigma\mathbf{k}'\sigma'}.
   \]

   The modified DOS can also be understood quantum mechanically. For simplicity we consider a one-dimensional system where the usual local Berry curvatures are absent. In this case, the gauge field $A_{k}(x)$ gives rise to a $k$- and $x$- dependent phase of the quasi-particle wave-function
        \begin{equation}
        \label{phase}
        u_{k}(x)\rightarrow u_{k}(x)e^{i\theta(k;x)},
        \end{equation}
        where $A_{k}(x)=d\theta(k;x)/dx$. We note that since $A_{k}(x)$ is a pure gauge, it does not have any effect on the local dynamics of the system. However, it may have a physical effect through modifying the boundary condition, i.e. a large gauge transformation.

        For ordinary periodic systems, the allowed momentum value $k$'s are quantized because of the boundary condition $e^{ikL}=e^{i0}=1$, which gives rise to the quantization condition $kL=2m\pi$, $L$=length of system and the spacing between allowed value of momentum is given by $\delta k=2\pi/L$. In the presence of nonzero $\theta(k;x)$ field, the boundary condition is modified to $kL+\theta(k;L)-\theta(k;0)=2m\pi$ and the spacing between allowed value of momentum is given by
        \begin{equation}
        \label{psa}
        \delta k(1+{1\over L}{d(\theta(k;L)-\theta(k;0))\over dk})={2\pi\over L}.
        \end{equation}
        We see that the DOS $g(k)$ in $k$ space is modified from $g(k)={L\over(2\pi)}$ to $g(k)={L\over(2\pi)}+z(k)$, where
         \begin{eqnarray}
           z(k) & = & {1\over2\pi}{d(\theta(k;L)-\theta(k;0))\over dk}  \\ \nonumber
           & = & {1\over2\pi}{d\over dk}\int_0^L dx {d\theta(k;x)\over dx}={1\over2\pi}\int_0^Ldx{d\over dk}A_k(x),
           \end{eqnarray}
         showing that we may identify $z(k,x)=(2\pi)^{-1}dA_{k}(x)/dk$ as the correction to {\em local} DOS of the system, which is exactly what we obtained from the $U(1)$ Fermi liquid state. The analysis can be extended to higher dimension systems easily in linear-response regime where $\partial(A_{\mathbf{k}}(\mathbf{r}))_i/\partial k_j (i\neq j)$ do not contribute\cite{Niu, Sun}.

          It is interesting to note that similar renormalization of DOS has been observed by Niu et al.\cite{Niu} in studying the semi-classical dynamics of electrons in periodic lattices when effective {\em real}- and $k$- space gauge fields (Berry curvatures) are present. let $\mathbf{A}_{\mathbf{k}}(\mathbf{r})$ and $\mathbf{B}_{\mathbf{k}}(\mathbf{r})$ be the Berry connection in real- and $k$- space, respectively. The crossed $r-k$ space Berry curvatures are defined by $\Omega^{rk}_{\mu\nu}=\partial_{r_{\mu}}B_{\nu}-\partial_{k_{\nu}}A_{\mu}$\cite{Niu, Sun}, etc. Renormalization in DOS appears when
          $\Omega^{rk}_{\mu\nu}\neq0$.
          The non-vanishing $k$-space derivative $\nabla_{\mathbf{k}}.\mathbf{A}_{\mathbf{k}}\neq0$ we discuss in this paper corresponds to a special case where $\mathbf{B}=0$ but $\Omega^{kr}_{\mu\mu}\neq0$. The crossed Berry-curvature is not a band structure effect as studied by Niu {\em et al.}, but an interaction effect driven by quasi-particle excitations. Our result in this paper suggests that interaction-induced crossed Berry curvature in $k-r$ phase space may provide an effective mechanism of generating non-Fermi liquid states. The $U(1)$ Fermi liquid state we propose in this paper is probably just one example in this class of states.

%\begin{thebibliography}
%\vspace{.5in}

%\end{thebibliography}

\end{document}